# Lévy Flights and Earthquakes

Sotolongo-Costa, O. [1]; Antoranz, J.C. [2]; Posadas, A. [3,4]; Vidal, F. [4, 5]; Vázquez, A. [1]

[1] *Departamento de Física Teórica. Universidad de la Habana. Cuba.*

[2] *Departamento de Física Matemática y Fluidos. UNED. Madrid. Spain*

[3] *Departamento de Física Aplicada. Universidad de Almería. Spain.*

[4] *Instituto Andaluz de Geofísica. Granada. Spain.*

[5] *Instituto Geográfico Nacional. Ministerio de Fomento. Madrid. Spain*

**Abstract**

Lévy flights representation is proposed to describe earthquake characteristics like the distribution of waiting times and position of hypocenters in a seismic region. Over 7500 microearthquakes and earthquakes from 1985 to 1994 were analyzed to test that its spatial and temporal distributions are such that can be described by a Lévy flight with anomalous diffusion (in this case in a subdiffusive regime). Earthquake behavior is well described through Lévy flights and Lévy distribution functions such as results show.
Key Words: Lévy flight, Self-organised criticality, seismicity.

**Introduction**

According to current plate tectonics, (Kanamori and Boschi, 1983) it is believed that lithospheric-plate motions, induced by mantlewide convection, create on-going plate-collisions and thus stresses which accumulate at the borders of the plates. This can be viewed as an incoming stress flux at the border of the system (we considered here a single lithospheric plate). When the plate, which is stressed, is deformed so strongly that the deformation exceeds a certain limit (which may change from place to place), a rupture occurs and an earthquake follows. An image of earthquake occurrence is that of sliding of two fractional brownian profiles. An earthquake occurs when there is an overlapping of the two profiles representing the two fault faces (Hallgass *et al*, 1996). The current research about earthquakes exhibits some important concepts recently introduced in Physics: Self-organised criticality (SOC). This SOC is perhaps one of the concepts recently most used in this field. Fractal geometry, hierarchical models, depinning and Lévy distributions are probably less used though intimately connected with SOC. Bak, Tang and Wiesenfeld (1987) proposed that spatio-temporal non-linear dynamical system, with quasi-static incoming and outcoming fluxes localized, for instance, at the borders, evolve spontaneously towards a stationary self-organised critical state. Neither length nor time scales others than those deduced from the size of the system and that of the elementary cell appear in the system. Subsequently it has been proposed that earthquakes may be an important natural phenomenon exhibiting SOC (Takayasu and Matsuzaki, 1988; Sornette and Sornette, 1989; Bak and Tang, 1989). These developments revived interest in much earlier work in which seismicity was modeled with blocks and springs. For instance, with a one dimension chain (Burridge and Knopoff, 1967), spawning a series of studies involving numerical models of block-spring systems of various types (Carlson and Langer, 1989a, 1989b; Nakanishi, 1990; Brown, Scholz and Rundle, 1991). Scholz (1991) has argued that earth's entire crust is in a state of self-organised criticality. Thus the crust is everywhere on the brink of failure. Sornette (1991) has given similar arguments. Earthquakes may be the cleanest and most direct example of a self-organised critical phenomenon in Nature, and this is widely recognized. Another important characteristic of earthquakes, *i.e.*, the migration of hypocenters and its description in terms of anomalous diffusion has not received enough attention, in spite of it is closely related with criticality. This is the main goal of our paper. The hypothesis of SOC for earthquakes leads to a power law for the temporal fluctuations for earthquake occurrence, which rationalize many observations. The Gutenberg-Richter law can be interpreted as a manifestation of the Self-organised critical behavior of the earth dynamics (Bak and Tang, 1988). Bak (1991) found that the powers may differ from different models, but there is also the distinct possibility, known from equilibrium critical phenomena as *universality* that power depends only on geometrical and topological features such as the spatial dimension. Several groups have suggested that self-organised criticality is a natural explanation for the Gutenberg-Richter law (Sornette and Sornette, 1989; Bak and Tang,

1989; Ito and Matzusaki, 1990; Correig et al, 1997). On the other hand, anomalous diffusion processes generally occur in disordered systems, *i.e.*, electronic conduction in amorphous semiconductors, atomic diffusion in glass like materials and others. The description of this phenomenon in terms of Lévy flights has shown to be adequate. (Vázquez, Sotolongo-Costa and Brouers, 1998). This kind of diffusion is usually modeled through a probability density of waiting times between successive steps in the walk, *Continuous Time Random Walks* (CTRW). The theory of CTRW has been extensively developed (Bouchaud and Georges, 1990 and references therein). The existence of a wide distribution of waiting times leads to a subdiffusive regime where the mean square displacement grows slower than time. As we will show in this paper, hypocenters are well described by a model of CTRW with a subdiffusive regime.

**Lévy Flight. Distribution of Hypocenters**

Arguments have been presented about the fractal character of the geographic distribution of hypocenters (Nakanishi *et al*, 1992) in the seismic region (we prefer to speak about the Levy distribution of hypocenters). This kind of distribution could be modeled by some type of *anomalous* diffusion determined by some dynamics based on *waiting times*. Earthquakes can be considered as a relaxation mechanism of the earth crust loaded with inhomogeneous stresses, which accumulate at lithospheric-plate borders. This inhomogeneity determines an irregular distribution of hypocenters. Once an earthquake occurs, the whole landscape of the stresses on the earth crust redistributes itself, and a new event will occur when the accumulated stresses surpass again the threshold somewhere else (incidentally, these arguments remind the *punctuated equilibrium* behavior in evolution and many other natural phenomena). The new place of occurrence will be considered here as the new position of a random walk, which has to wait for a time $\tau_w$ on each site before the next jump. Once an earthquake has occurred somewhere, we can assume that the random walk has to wait until the redistribution of stresses leads to a new earthquake somewhere else, at a distance x from the place of the former seism. This jump occurs suddenly, so that the waiting time $\tau_x$ for the walker to go from one point to another is much less than the waiting time $\tau_w$ in the place where the last earthquake has occurred. The waiting time is a random variable distributed according to a given law $p(\tau_w)$. We also assume that the waiting time is not correlated to the length of the jump x, distributed as p(x). The distributions $p(\tau_w)$ and p(x) for a given seismic region should differ. Indeed, since earthquakes occur mainly in some limited regions (seismic regions), a size effect is imposed to the geographic distribution of hypocenters. Then $p(x)$ must have finite variance. Assuming that the tail of p(x) is described by a power law ($p(x) \approx x^{-a_x}$) then $a_x \geq 2$ to ensure finite variance. If we limit our analysis to a seismic region then we can imagine the hypocenter as a random walk confined in a given region with a wide distribution of waiting times. We adopt the CTRW model to describe the migration of earthquakes in a given seismic region. This standpoint is supported by the representation of earthquakes as the slipping between asperities, where displacements between blocks of a fault occur leading to the seism. This model is a good tool to represent the migration of hypocenters in a seismic region as a problem of the diffusion of a random walk in a comb-like structure. So, CTRW analysis is directly applicable and has well known results for the distribution of waiting times and mean square displacement (Buochaud and Georges, 1990). The diffusion process is characterized by the scattering function $F(k,t)$, the Fourier transform of the diffusion front. Properties like the mean square displacement can be derived from this function, *i.e.*:

$$<x^2> = \partial^2 F / \partial k^2 \big|_{k=0} \quad (1)$$

Let $N$ be the number of steps performed by a walker during time $t$. $N$ is, in general, a random variable, which depends on the duration of the jumps and waiting times. The scattering function can thus be expressed as a sum over all possible jumps during time:

$$F(k,t) = \int F(k,N) P(N,t) dN \quad (2)$$

where $F(k,N)$ is the scattering function of the same problem, but considering regular duration of the jumps and no waiting time, and $P(N,T)$ stands for the probability

distribution of $N$ jumps at a fixed time $t$. The total displacement after $N$ steps is given by:

$$X_N = \sum_{i=1}^{N} x_i \tag{3}$$

In the right hand side we have a sum of mutually independent random variables with the common distribution p(x), with zero mean. The limit distribution for large $N$ will be a stable Lévy distribution (Feller, 1966; Gnedenko, 1954), i.e.:

$$X_N \approx l^* N^{1/\alpha_x} u \tag{4}$$

where the symbol $\approx$ denotes that random variables in both sides have the same distribution and $u$ follows the symmetric Lévy distribution $L_{\alpha,0}(u)$. If p(x) has finite variance then $l^* = \sigma$ and $\alpha_x = 2$, while if p(x) ~ $l_0^\alpha |x|^{-1-\mu}$ with 0<μ<2 then $l^* \sim l_0$ and $\alpha_x = \mu$. The canonical (Fourier transform) representation of Lévy stable laws is (for $\alpha \neq 1$):

$$FT[L_{\alpha,\beta}](k) = \exp\left[-|k|^\alpha (1 + \frac{k}{|k|}) i\beta \tan(\alpha\pi/2)\right] \tag{5}$$

where α and β are real numbers defined in the intervals 0<α≤2 and -1≤β≤1. The case α=2 and β=0 corresponds with the Gaussian distribution, which decays faster than any power law for large arguments. On the contrary, all Lévy distributions, except the Gaussian, have the asymptotic behavior for $u \gg 1$ (Feller, 1966; Gnedenko, 1954):

$$L_{\alpha\beta}(u) \propto u^{-1-\alpha} \tag{6}$$

Then, from equations (4) and (5), it follows that:

$$F(k, N) = \exp\left[-N(l^*|k|)^{\alpha_x}\right] \tag{7}$$

On the other hand, the number of steps after time $t$ is given by (assuming $\tau_w \gg \tau_x$):

$$t = \sum_{i=i}^{N} t_{wi} \tag{8}$$

where $\tau_{wi}$ are the waiting times between earthquakes. In the right hand side of equation (8) we have a sum of independent random variables, with distribution p($\tau_w$). The limit

distributions for large $N$ will follow Lévy distributions (Feller, 1966; Gnedenko, 1954):

$$t \approx \mathbf{t}N + \mathbf{t}_w^* N^{1/a_w} u \qquad (9)$$

where $u$ follows the Lévy distribution $L_{a_1}(u)$. The first term in the right hand side appears only if $p(\tau w)$ has finite mean, and $\tau$ is given by the sum of the finite means. If $p(\tau_w)$ has finite variance ( then $\mathbf{a}_w^* = \sigma$ and $\alpha_w = 2$, while if $p(\tau_w) \sim \tau_0^{\alpha w} \tau_w^{-1-\mu}$ ($0<\mu<2$) then $\tau_w \sim \tau_0$ and $\alpha_w = \mu$. Let us assume that the distribution of waiting times obeys a very wide distribution function so that $\alpha_w$ can be assumed to be $\alpha_w = 1$ and the first term in equation (9) does not appear. The restriction of earthquakes to some regions of the earth crust is an argument to support the hypothesis that $\alpha_x >> \alpha_w$. From equations (2), (7) and (9) it follows that:

$$F(k,t) = \int L_{a_w,1} \exp\left[-N(kl^*)^{a_w}\right] du \qquad (10)$$

where the functional dependence of $N$ with $u$ and $t$ is determined from equation (9). If, as assumed, the distribution of waiting times is wide, the second term of the r.h.s. is dominant and $N = (t/u\mathbf{t}_w)\mathbf{a}_w$. To use this result and equation (1) in (10) to obtain the mean square displacement, we must perform a realistic evaluation taking into account that the total displacement at time $t$ cannot be larger than $vt$, where v is the velocity of displacement of the hypocenter and therefore there is a cut-off $k_c \sim t^{-1}$ for small values of $k$. Thus we evaluate equation (1) in $k = k_c$ instead of in $k = 0$. We obtain:

$$<x^2> \sim t^{a_w - a_x + 2} \qquad (11)$$

As we have supposed that the geographic distribution of hypocenters has finite variance, we put, in a rough approximation, $\alpha_x \cong 2$, obtaining:

$$<x^2> \sim t^{a_w} \qquad (12)$$

This corresponds to a subdiffusive behavior since the wide character of waiting times implies small values for $\alpha_w$. To gain insight in the allowed values for $\alpha_w$ its is helpful to relate the nature of the rough profile of faults with the known problem of CTRW in

comb-like structures. In that model the waiting time distribution function is (Bouchaud and Georges, 1990):

$$\Psi(t) \sim t^{(1+a_w)} \sim t^{1.5} \tag{13}$$

If diffusion occurs in a comb-like structure as that evoked by the fault profile, this then implies $<x^2> \sim t^{0.5}$ for the mean square displacement. To check the validity of this assumption in the next section we will analyze the diffusion of hypocenters in the Central Betics zone.

**Data from South Spain Earthquakes**

The Andalusian Seismic Network (Alguacil, 1986) (located in Southern Spain) includes 18 stations for microearthquake detection, 10 accelerographs for strong and weak motions and 6 Broad-Band sets. This network, belonging to the Andalusian Geophysics Institute, provide us a wide seismic catalogue that involve thousands of earthquakes and microearthquakes. The location feasibility of the network allow us to have high precision hypocenter determination (Serrano, 1999). The area under study has a high activity microearthquakes with hypocenters shallower than 20 km. However also intermediate and deep seismic activity is detected (Vidal, 1986; Morales et al, 1997). From the point of view of seismic activity, Southern Spain is the region with the highest hazard level in Spain due to it is located in the interaction zone between the Euroasian and African plates. The area under study is situated in the central part of the Betic Cordilleras (Southern Spain) (figure 1). The structure of the crust is characterized by a rougly flat Moho (~38 km) with a suddenly change in the Moho depth in the transition to the Alboran sea domain (Galindo-Zaldivar et al, 1997; Serrano, 1999) .The faulting present in the zone created a set of blocks that are structured at different levels that allow independent movements of them. There are also compressive and extensional coeval deformations (Galindo-Zaldivar et al, 1999; Morales et al, 1999). These features fit into a general compressive (Morales et al, 1999) framework, which produces contemporary extensional and compressive deformations. The seismogenetic areas are concentrated in three fracture systems having N10-30E, N30-60W and N70-100E directions and all the fracture systems are embedded in the Betic Area (Vidal, 1986; Peña et al, 1991; Posadas et al, 1993a; Posadas et al, 1993b).

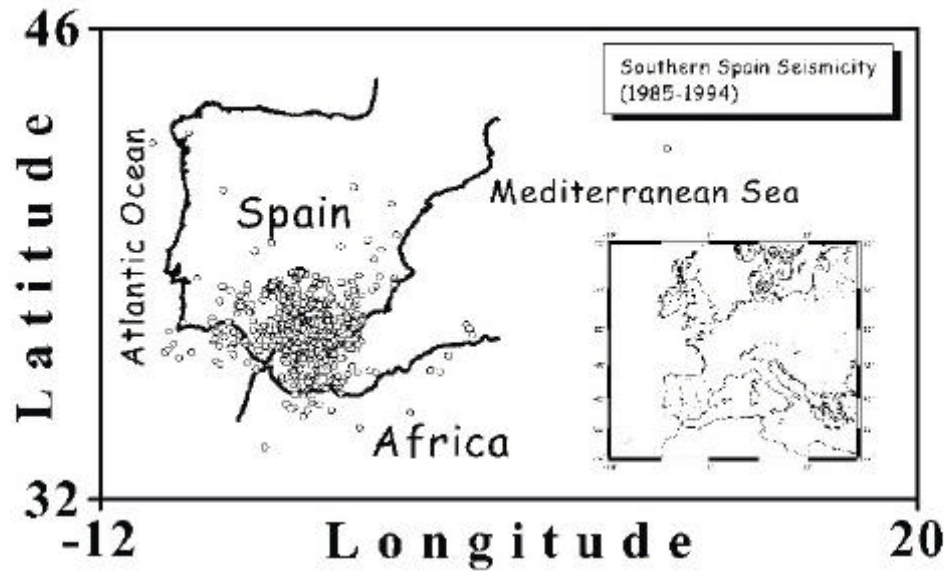

*Figure 1. The area under study is located in the interaction zone between Europe and Africa plates. More than 7500 epicenters were studied. Longitude and Latitude are, respectively, East Longitude and North Latitude. Data belonging to the Andalusian Institute of Geophysics are recollected by the Andalusian Seismic Network.*

**Results and conclusions**

Over 7500 microearthquakes and earthquakes from 1985 to 1994 (figure 1) were analyzed to test that its spatial and temporal distributions are such that can be described by a Lévy flight with anomalous diffusion (in this case in a subdiffusive regime). Figure 2 shows the result after to apply Lévy Flight Model to the present data. Normalized distance *vs* time were used to depict units: $D$ is the average distance between earthquakes and $\tau$ was chosen to be the average time between two consecutively earthquakes. It has been found that data can be modeled with a potential law $y = a + bx^c$, with parameters $a \cong -0.4$, $b \cong 2$ and $c \cong 0.5$, indicating its subdiffusive behavior. The correlation factor was $R^2 = 0.9974$ although it necessary to understand that because of the cumulative curves include non-negative variables, such procedure overestimate $R^2$ value. High correlation factor shows that we have found a very good confirmation of equation (12) with $\alpha_w$ predicted by equation (13). Correspondingly, figure 3 shows the waiting time distribution function, *i.e.*, the normalized time distribution between earthquakes for all the former processed events. The graph shows a logarithmic plot with a slope around 1.7. This fact corroborates in a satisfactory measure our standpoint.

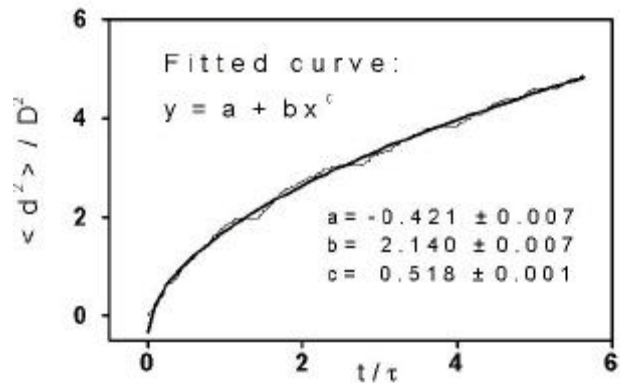

*Figure 2. Applying Lévy Flight Model to the present data, normalized distance versus time were used to despict units. Thin line represents results from actual data; wide line is the fitted curve. It has been found that data can be modeled with a potential law $y = a + bx^c$, with $c \approx 0.5$, indicating its subdiffusive behavior.*

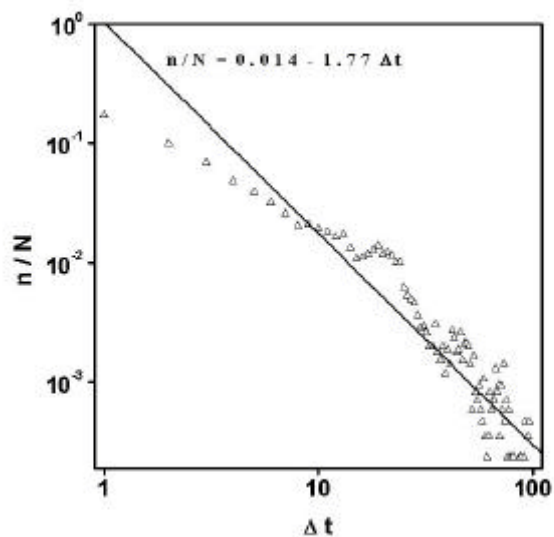

*Figure 3. The waiting time distribution function, i.e., the normalized time distribution between earthquakes for all the former processed events. The graph shows a logarithmic plot with a slope around 1.7.*

The concept of SOC provides a very stimulating framework within which to tackle the problem of defining and understanding the mechanics of the Lithosphere, *i.e.*, the connection between rupture at short time scale to that at large time scales and the geometric characteristics (scaling, Lévy distributions, etc.) essentially linked with it. We have found an anomalous diffusive behavior (subdiffusive behavior) in the description of hypocenter migration. This indicates that the system (crust in the Central Betic Area) follows a non-equilibrium dynamics. The migration of earthquakes can be described as the diffusion of a walker in a comb-like structure, *i.e.*, it can be described with the CTRW model. A Lévy flight description was used to characterize spatial and temporal distribution of earthquakes. Moreover, the exponent of the model for the mean square displacement has been quantitatively determined from observations and its value is 0.5184 with correlation factor equal to 0.9974. This value determines the temporal occurrence of earthquakes, as was here shown applying the results of CTRW model.

**Acknowledgements**

This work was partially supported by the CICYT projects AMB97-1113-CO2-02, AMB99-1015-CO2-02, the DGESIC project AMB1999-0129 (Almería University, Spain) and the Alma Mater prize given by the Havana University.

**References**

Alguacil, G.; (1986); Los instrumentos de una red telemétrica para microterremotos. La Red Sísmica de la Universidad de Granada. PhD. Thesis. 228 pp. University of Granada.

Bak, P.; (1991); Self-organised criticality and the perception of large events. In " Spontaneous formation of space-time structures and criticality". Riste, T; Sherrintong, D. Eds. Kluwer Academic Publishers.

Bak, P.; Tang, C.; Wiesenfeld, K.; (1987);" Self-organised criticality: an explanation of 1/f noise". Phys. Rev. Lett., 59, 381-384.

Bak, P.; Tang, C.; Wiesenfeld, K.; (1988); Self-organised criticality. Phys. Rev. A, 38, 364-374.

Bak, P.; Tang, C.; (1989); Earthquakes as a Self-organised critical phenomenon. J. Geophys. Res., 94, 15635-15637.

Bouchaud, J. P. and Georges, A. (1990); Anomalous diffusion in disordered media: Statistical mechanisms, models and physical applications. Physics Reports 195 Nos. 4 & 5, 127-293.

Brown, S.; Scholz, C.; Rundle, J.; (1991); A simplified spring-block model of earthquakes. Geophys. Res. Lett., 18, 215-218.

Burridge, R.; Knopoff, L.; (1967); Model and theoretical seismicity. Bull. Seism. Soc. Am, 57, 341-371.

Carlson, J.M.; Langer, J.S.; (1989a); Properties of earthquakes generated by fault dynamics. Phys. Rev. Lett., 62, 2632-2635.

Carlson, J.M.; Langer, J.S.; (1989b); Mechanical model of an earthquake fault. Phys. Rev. A, 40, 6470-6484.

Correig, A.; Urquizú, M.; Vila, J.; (1997); Aftershocks series of event February 18, 1996: an interpretation in terms of Self-organised criticality. J. Geophys. Res. 102, B12, 27407-27420.

Feller, W.; (1966); An Introduction to Probability Theory and Applications. Vol. 2. Wiley, New York, 1966.

Galindo-Zaldivar, J., A. Jabaloy, I. Serrano, J. Morales, F. Gonzalez-Lodeiro & F. Torcal (1999). Recent and present-day stresses in the Granada basin (Betic Cordilleras: Example of a late Miocene-present-day extensional basin in a convergent plate boundary. Tectonics, vol 18, 686-702.

Galindo-Zaldívar, J., Jabaloy, A., González-Lodeiro, F., and Aldaya, F., 1997, Crustal structure of the central sector of the Betic Cordillera (SE Spain): Tectonics, v. 16, p. 18-37.

Gnedenko, B.V. and A. N. Kolmogorov; (1954); Limit distributions for sums of Independent Random Variables. Addison Wesley Reading, MA.

Hallgass, R., Loreto, V., Mazzella, O., Paladin, G., Pietronero, L.; (1996); Earthquake statistics and fractal faults. cond-mat/9606153.

Ito, K.; Matsuzaki, M.; (1990); Earthquakes as Self-organised critical phenomena. J. Geophys. Res. 95, B5, 6853-6860.

Kagan, Y.; Knopoff, L.; (1980); The spatial distribution of earthquakes: the two-point correlation function. Geophys. J. R. Astron. Soc. 62, 303-320.

Kanamori, H.; Boschi, E. (Editors); (1983); Earthquakes, observation, theory and interpretation. North-Holland, Amsterdam.

Morales, J., I. Serrano, F. Vidal & F. Torcal (1997).The depth of the earthquake activity in the Central Betics (Southern Spain).Geophys. Res. Lett. 24:3289-3292.

Morales, J., I. Serrano, A. Jabaloy, J. Galindo-Zaldivar, D. Zhao, F. Torcal, F. Vidal & F. Gonzalez-Lodeiro (1999). Active continental subduction beneath the Betic Cordillera and Alborán Sea. Geology Vol. 27: 735-738.

Nakanishi, H.; (1990); Cellular automation model of earthquakes with deterministic dynamics. Phys. Rev. A, 41, 7086-7089.

Nakanishi, H., Sahimi, M., Robertson, M. C., Sammis, C. C., Rintoul M. D.; (1992); Fractal properties of the distribution of earthquake hypocenters. J. de Physique, 3, 733-739.

Peña, J.; Vidal, F.; Posadas, A.; Morales, J.; Alguacil, G.; De Miguel, F.; Ibáñez, J.; Romacho, M.; López-Linares, A.; (1993); Space clustering properties of the Betic-Alboran earthquakes in the period 1962-1989. Tectonophysics, 221, 125-134.

Posadas, A.M.; F. Vidal; F. De Miguel; G. Alguacil; J. Peña; J.M. Ibañez; J. Morales; (1993a); "Spatial-temporal analysis of a seismic series using the Principal Components Method. The Antequera Series (Spain), 1989. J. Geophys. Res., 98, 1923-1932.

Posadas, A.M.; F. Vidal; J. Morales; J.A. Peña; J. Ibañez; F. Luzon; (1993b); "Spatial and temporal analysis of a seismic series using a new version of three point method. Application to Antequera (Spain) 1989 earthquakes". Physics of the Earth and Planetary Interiors, 80, (1993), pp. 159-168.

Scholz, C.; (1991); Earthquakes and faulting: Self-organised critical phenomena with characteristic dimension. In Spontaneous Formation of space-time Structures and Criticality. Riste, T. And Sherrington, D. Eds. Kluwer Academic Publishers. The Netherlands.

Serrano, I. (1999); Distribución espacial de la sismicidad en las Cordilleras Béticas-Mar de Alborán. PhD thesis. Universidad de Granada. 231 pp.

Sornette, D.; (1991); Self-organised Criticality in Plate Tectonics. In Spontaneous formation of space-time structures and criticality. Riste, T; Sherrintong, D. Eds. Kluwer Academic Publishers.

Sornette, A.; Sornette, S.; (1989); Self-organised criticality and earthquakes. Europhys. Lett. 9, 197-202.

Takayasu, H.; Matsuzaki, M.; (1988); Dynamical phase transition in threshold elements. Phys. Lett. A, 131, 244-247.

Vazquez, A., Sotolongo-Costa, O., Brouers, F. (1998); Diffusion regimes in Lévy flights with trapping. Physica A 264 424-431.

Vidal, F.; (1986); Sismotectónica de la región Béticas-Mar de Alborán. PhD Thesis. Universidad de Granada. 450 pp.